# Design and test of SX-FEL Cavity BPM

YUAN Ren-Xian ZHOU Wei-Min Chen Zhi-Chu YU Lu-Yang WANG Bao-Pen LENG Yong-Bin

Shanghai Institute of applied physics, Shanghai 201800, China

**Abstract:** This paper reports the design and cold test of the cavity beam position monitor (CBPM) for SX-FEL to fulfill the requirement of beam position measurement resolution of less than 1μm, even 0.1μm. The CBPM was optimized by using a coupling slot to damp the $TM_{010}$ mode in the output signal. The isolation of $TM_{010}$ mode is about 117dB, and the shunt impedance is about 200Ω@4.65GHz with the quality factor 80 from MAFIA simulation and test result. A special antenna was designed to load power for reducing excitation of other modes in the cavity. The resulting output power of $TM_{110}$ mode was about 90mV/mm when the source was 6dBm, and the accomplishable minimum voltage was about 200μV. The resolution of the CBPM was about 0.1μm from the linear fitting result based on the cold test.

**Introduction**

The Soft X-ray Free-Electron Laser Project in Shanghai requires a very precise beam position measurement with the resolution of about 0.2μm@1nC. Among various types of BPMs, CBPM is the indubitable monitor to achieve the resolution. Usually the Quality factor is a very important parameter for the CBPM. When the Q factor is higher than 1000, it means the coupling structure has a low efficiency. Then, the output signal will have a small amplitude, and a long last time simultaneously. If the Q factor is less than 100, it means the coupling structure has a very high efficiency. The output signal will have a large amplitude and a short time at the same time. There are many CBPM's experiment reports with nanometer level resolution[1-3]. Considering the bunch-by-bunch beam position diagnostic, the high Q cavity will cause signals mixing and can't give an adequate position resolution. So the CBPM in SX-FEL adopts only the low Q scheme. After considering the vacuum pipe radius and to avoid the dark current from the accelerating RF system, the working frequency of CBPM was chosen to be 4.65GHz.

**Design of the CBPM structure**

For a cylindrical pill-box cavity, a series of TM modes and TE modes will be excited. Because the beam source is along the z-axis, the bunch will not lose any energy to the transverse E-fields of the TE modes. Hence only TM modes will be excited. The amplitudes of the TM modes are determined by the bunch's lost energy. Usually the base mode $TM_{010}$ will get the most lost energy because its E-field is the maximum on the z-axis, and the position mode $TM_{110}$ will get a very small lost energy for its E-field is about zero when bunch is very close to the z-axis. It is very important to design an appropriate coupling structure which can export $TM_{110}$ mode very well and damp most of the power from $TM_{010}$ mode at the same time [3-4]. Usually the amplitude ratio of $TM_{010}$ mode and $TM_{110}$ mode can be written as:

$$K_r = \frac{(R/Q)_{010}^{0.5}}{(R/Q)_{110}^{0.5}} = \sqrt{\frac{2f_{110}}{f_{010}}} \cdot \frac{J_0(rk_{010})J_0(R_c k_{011})}{J_1(rk_{011})J_1(R_c k_{010})} \quad (1)$$

Where $J_0$, $J_1$ are the Bessel functions, and $R_c$ is the radius of the cavity, and r is the offset of the bunch from the center of the cavity, and $k_{010}$, $k_{011}$ are the wave numbers of $TM_{010}$, $TM_{011}$ modes. Usually, the ratio $K_r$ will be about 10000 when beam offset is 1μm. It can be seen that a little power leakage from $TM_{010}$ mode will cause a large noise which will reduce the position resolution of CBPM. If the resolution about 0.1μm is needed, it means that the amplitude of the leaking power from $TM_{010}$ mode should be less than the amplitude of $TM_{110}$ mode when the bunch's offset is only 0.1μm. It means that the $TM_{010}$ mode should be damped 100dB at least.

For rejecting $TM_{010}$ mode, the coupling structure was designed as a slot along the z-axis, and the energy of $TM_{011}$ mode was exported as magnetic field coupling. The structure of the CBPM was designed as Fig.1. The amplitude and its spectrum of the output signal were simulated in Mafia and shown in Figs.2 and 3 when the bunch's offset was only 0.5mm.

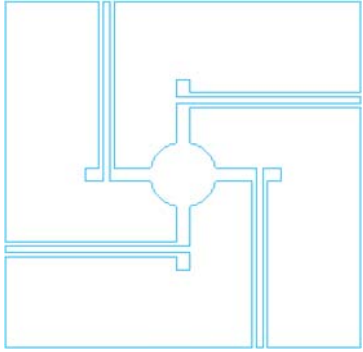

Fig.1. Structure of the prototype CBPM

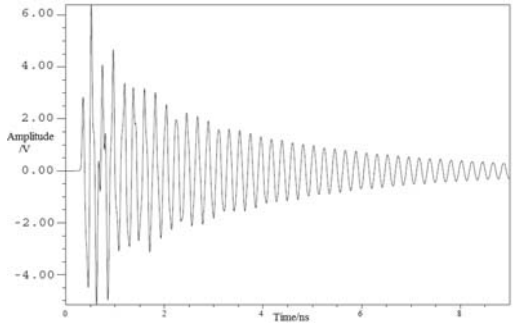

Fig.2. Output voltage of the CBPM in simulation with beam charge 1nC and offset 0.5mm

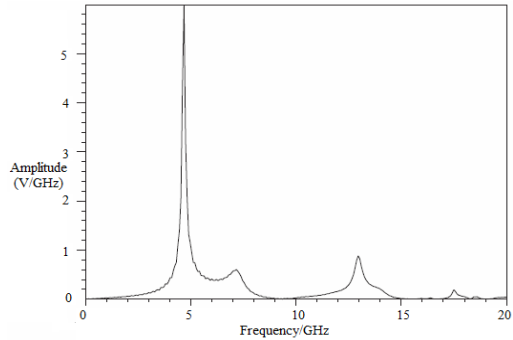

Fig.3. Spectrum of the CBPM when simulation

From the simulation, the eigenfrequencies of $TM_{010}$ and $TM_{110}$ modes were 3.6GHz and 4.65GHz, respectively. It can be seen that the $TM_{010}$ mode is damped about 117dB in Fig.3. So the resolution of the CBPM is less than 0.01μm if only considering the noise from $TM_{010}$ mode's leakage. It can be also seen that the load Q factor is about 80, it means that the magnetic coupling structure has very high efficiency. From the spectrum, the amplitude at the exited 50Ω impedance can be evaluated as:

$$V(t) = V(f) \cdot BW \cdot \exp(-t/Q_L T_0) \quad (2)$$

Where V(f) is the amplitude of $TM_{110}$ mode around its eigenfrequency, and BW is the bandwidth of the CBPM's RF front-end. So the received peak voltage of ADC is about 0.48mV/μm when the BW is supposed to be 20MHz, and V(f) is about 24μV/MHz*μm. Meanwhile, the RMS thermal noise is about 0.92μV/MHz, so the total noise is about 0.004mV. It means that the beam position voltage of the CBPM will be equal to the thermal noise when the offset of beam is about 9nm.

**Cold test of SX-FEL CBPM**

After manufacturing a prototype CBPM, a calibration platform was built in Feb, 2012, then the prototype CBPM was tested. The CBPM was laid on a 2-dimisional moving platform with the minimum step of 10nm, and this platform was also laid on a manual transverse platform with 20μm/step. All the CBPM and the transverse platforms were laid on a manual lifter to adjust the center of the CBPM.

Usually there are two ways to load the power into the CBPM to simulate the beam excitation. One is the load wire [4-5], and another is antenna. It is very hard to load the power into the CBPM because the impedance of the load wire could not match. On the other hand, the vibration problem on a long wire is quite serious. It is difficult to get a calibrated resolution lower than 10μm using load wire in cold test.

The difference between the beam field and the antenna field is very large. It is well known that the beam field is translational symmetry on z-axis for high energy electron beam, but it is very hard to produce an antenna field with translational symmetry. So after making several antennas, some TE modes were observed as shown in Fig.4. The TE modes whose eigenfrequencies are close to the CBPM's working frequency will induct additional noise on the output port of the CBPM. The TE modes' effect can be evaluated using the accomplishable minimum voltage of CBPM as shown in Fig.5, where the minimum voltage was 1.1mV when the sensitivity of the CBPM was about 0.09mV/μm and the source power was 6dBm.

After many attempts, a special antenna was made which was loaded on both ports. With this antenna structure, the TE modes will be hardly excited and the loaded power has very high efficiency as seen in Fig.6. The accomplishable minimum voltage was about 200μV as shown in Fig.7 when the source power was also 6dBm, and the sensitivity kept at 0.09mV/μm too.

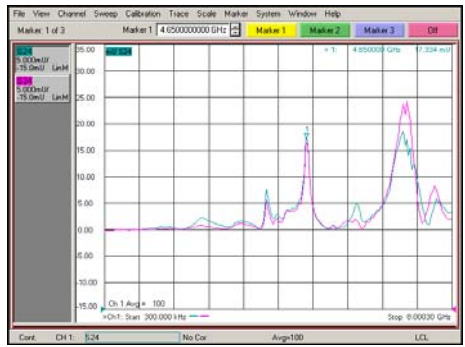

Fig.4. S-parameter test when having strong TE modes coupling

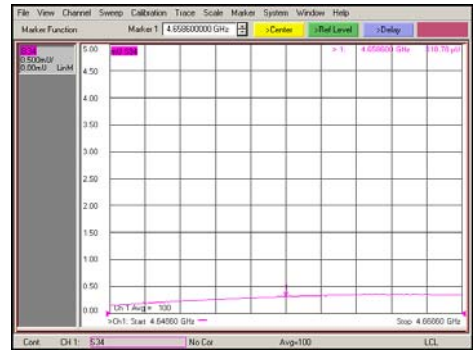

Fig.7 The accomplishable minimum voltage after improving antenna

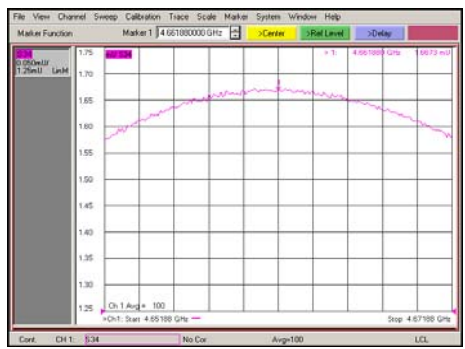

Fig.5. The accomplishable minimum voltage when using common antenna

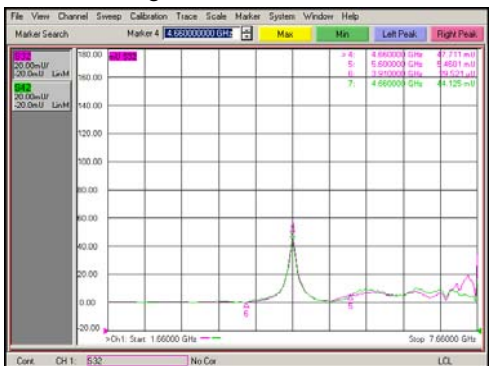

Fig.6 S-parameter test when having small TE modes coupling

The cold test results of the prototype CBPM were shown in Figs.8 and 9. Fig.8 was the test result when the antenna was near the center of the CBPM with the accomplishable minimum voltage as low as 40μV, but unfortunately this antenna was damaged later. Fig.9 was also the relations between the output voltages of the CBPM and the positions of antenna when the antenna had different offsets. There were some differences, although small, in all 4 test groups. The platform had to be adjusted by hand to get the difference offsets, so there would be some additional differences, such as axis angle between the CBPM and antenna in spite of the offsets. From the cold test, the resolution of the CBPM in cold test state was about 0.1μm when the CBPM's sensitivity was about 0.09mV/μm. So if the CBPM's sensitivity is about 0.48mV/μm with beam charge 1nC, the resolution of the CBPM is expected to 0.02μm

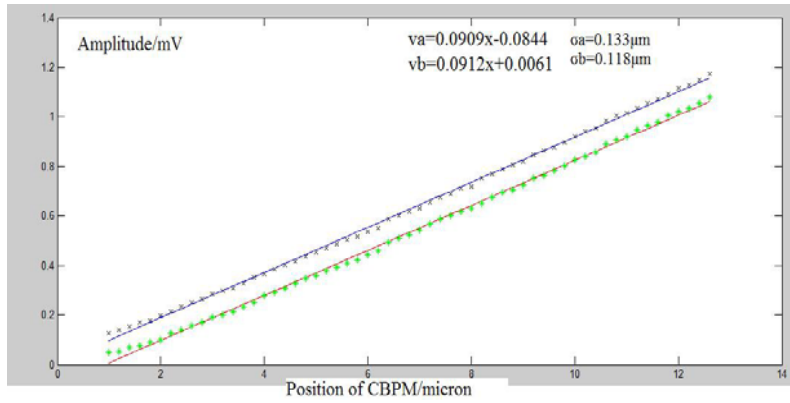

Fig.8 Resolutions of CBPM in cold test when the antenna was near the center of the CBPM after improving the antenna when the accomplishable minimum voltage was even about 40μV

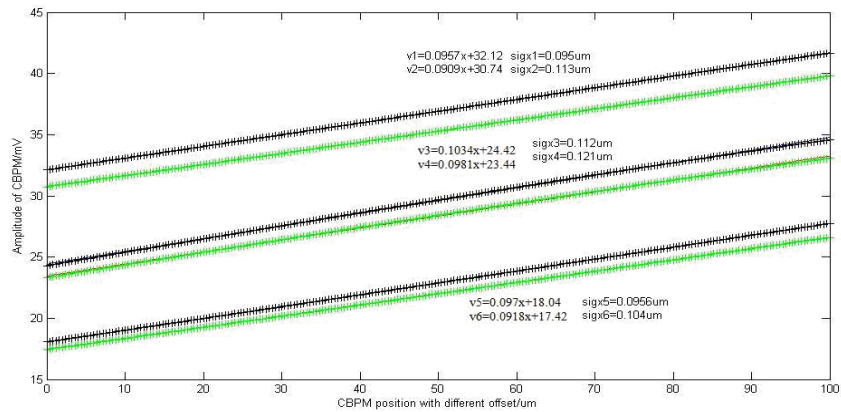

Fig.9. Resolutions of CBPM in cold test when the antenna had different offsets

**Conclusion**

A prototype CBPM was designed and tested for Shanghai SX-FEL with the help of a special designed antenna and a high accuracy platform. The resolution of the CBPM was got to be 0.1μm in cold test. A formal in-vacuum CBPM is being manufactured and will be tested soon. Considering the cold test was only based on net analyzer, the CBPM will be tested after obtaining the customized RF front-end and the digital signal processor.